\documentstyle[prc,twocolumn,aps,epsf,eqsecnum,ifthen,epsfig]{revtex}

\newcommand{\nc}{\newcommand}
\nc{\pt}{p_{\rm T}}
\nc{\se}{\section}
\nc{\suse}{\subsection}
\nc{\beq}[1]{\begin{equation}\label{#1}}
\nc{\eeq}{\end{equation}}
\nc{\bea}[1]{\begin{eqnarray}\label{#1}}
\nc{\eea}{\end{eqnarray}}
\nc{\bce}{\begin{center}}
\nc{\ece}{\end{center}}
\nc{\bit}{\begin{itemize}}
\nc{\eit}{\end{itemize}}
\nc{\bmp}{\begin{minipage}}
\nc{\emp}{\end{minipage}}

\nc{\la}{\langle}       
\nc{\lla}{\left \langle}
\nc{\ra}{\rangle}       
\nc{\rra}{\right \rangle}

\newcommand{\eg}{{\it e.g.}}
\newcommand{\ie}{{\it i.e.}}

\newcommand{\prevc}[3]{Phys. Rev. {\bf C#1}, #3 (#2)}
\newcommand{\prevd}[3]{Phys. Rev. {\bf D#1}, #3 (#2)}
\newcommand{\prevl}[3]{Phys. Rev. Lett.\ {\bf #1}, #3 (#2)}

\newcommand{\plb}[3]{Phys. Lett. {\bf B#1}, #3 (#2)}
\newcommand{\npa}[3]{Nucl. Phys. {\bf A#1}, #3 (#2)}
\newcommand{\npb}[3]{Nucl. Phys. {\bf B#1}, #3 (#2)}

\begin{document}

\twocolumn[\hsize\textwidth\columnwidth\hsize\csname @twocolumnfalse\endcsname

\title{Transverse Flow and Hadro-Chemistry 
in Au+Au Collisions at $\sqrt{s_{\rm NN}} = 200$~GeV}   
\author{Peter F.~Kolb$^a$ and Ralf Rapp$^b$}
\address{$^a$Department of Physics and Astronomy, 
             SUNY,
             Stony Brook, NY 11794-3800, USA\\
         $^b$NORDITA, Blegdamsvej 17, DK-2100 Copenhagen {\O}, Denmark
         }
\date{April 28, 2003}

\maketitle


\begin{abstract}

We present a hydrodynamic assessment of preliminary particle spectra
observed in Au+Au collisions at $\sqrt{s_{\rm NN}}=200$~GeV. 
The hadronic part of the underlying equation of state is based on explicit 
conservation of (measured) particle ratios throughout the resonance gas 
stage after chemical freezeout by employing  
chemical potentials for stable mesons, nucleons and antinucleons. 
We find that under these conditions the data 
(in particular the proton spectra) favor a low freezeout temperature of 
around $\sim$~100~MeV. 
Furthermore we show that through inclusion of a moderate pre-hydrodynamic 
transverse flow field the shape of the spectra improves with respect to 
the data.
The effect of the initial transverse boost on elliptic flow 
and the freezeout geometry of the system is also discussed.
\end{abstract}
\vspace{0.2in}
]
\begin{narrowtext}
\newpage

\section{Introduction}  
\label{sec_intro}
During its second year of operation, RHIC 
(the Relativistic Heavy Ion Collider at Brookhaven National Laboratory) 
has collided $^{197}$Au nuclei at center-of-mass ($CM$) energies
of 200~GeV per nucleon pair to create strong interaction 
matter at high energy densities in the laboratory.  
To identify signals of a possible phase transition from low-energy nuclear
to deconfined quark-gluon matter, 
a large amount of data was analyzed and recently presented for the first 
time~\cite{QM02}.
In the present article we investigate single particle spectra of various 
hadronic species within a hydrodynamic framework for the reaction dynamics, 
which assumes rapid thermalization in the reaction volume and a subsequent 
expansion according to the conservation of energy, momentum, entropy and 
baryon-number (for details of the approach, which resides on explicit 
longitudinal boost invariance, cf.~Ref.~\cite{KSH00}).
At $CM$ energies of 130~$A$GeV, the successful description   
of observed single-particle transverse momentum ($\pt$) spectra 
and their azimuthal modulation in non-central collisions have 
validated this approach down to decoupling temperatures of 
$\sim$~130~MeV, at which hadronic interactions have been assumed 
to cease instantaneously~\cite{hydrov2}.
As an alternative to entirely hydrodynamic simulations, especially
for the late, more dilute stages in a heavy-ion collision,  
hybrid models have been developed~\cite{BD00,SBD01,TLS01} 
which treat the hadronic phase in sequential-scattering models, 
propagating hadrons individually.
While the momentum-space observables are in good agreement with 
experiments at RHIC in both descriptions, the freezeout geometry
persists to be inconsistent with the data in either approach.
Hydrodynamic evolutions appear to be too 
long-lived but too small in radial extent~\cite{HK02}, whereas 
hybrid calculations produce an emission cloud which appears to be 
too large~\cite{BD00,SBD01}.
Concerning global particle production, it was soon realized~\cite{BMMRS01} 
that, also at RHIC energies, measured hadron ratios reflect a chemical 
composition of the fireball which corresponds to a temperature close to 
the expected QCD phase boundary, $T_{chem}\simeq$~170-180~MeV~$\simeq T_c$.
Thus, in a thermodynamic description of the cooling process from chemical 
to thermal freezeout, the conservation of the relative hadronic abundances 
requires the introduction of (effective) chemical 
potentials~\cite{KR90,BGGL92,HS98,AGHS01,Rapp02,Teaney02} 
for species that are stable on the scale of typical fireball lifetimes.  
In particular, it was pointed out in Ref.~\cite{Rapp02} that the conservation
of antibaryons plays an important role at collider energies.
Despite their large annihilation cross sections, their finally observed
abundance is in complete agreement with chemical-freezeout systematics 
(for a possible microscopic explanation of this fact, based on multi-meson 
fusion reactions to maintain detailed balance, cf.~Ref.~\cite{RS01}). 
This implies the build-up of large antibaryon chemical potentials,  
$\mu_{\bar{\rm N}}^{\rm eff}$, defined via  
$\mu_{\bar{\rm N}}=-\mu_{\rm N}+\mu_{\bar{\rm N}}^{\rm eff}$.
Towards thermal freezeout this, in turn, entails rather large baryon 
chemical potentials ($\mu_N \simeq 350$~MeV), and is at the origin    
of appreciable pion chemical potentials ($\mu_\pi\simeq$~80-100~MeV). 
The influence of chemical potentials on the hydrodynamic evolution and 
resulting observables has been investigated in Ref.~\cite{HT02} for 
$CM$ energies of 130~$A$GeV. 
In addition to conserving $\pi$-, $K$-, $\eta$- and $\eta'$-numbers, 
we here explicitly distinguish chemical potentials 
of baryons and antibaryons along the lines of Ref.~\cite{Rapp02} 
to correctly account for the finite net-baryon density at  
full RHIC energy (200~$A$GeV). 
For consistency with previous analyses~\cite{KSH00,hydrov2,KSH99,KHHET01} 
we assume  a phase transition from quark-gluon to hadron matter at  
$T_c=165$~MeV with a latent heat of $e_{{\rm lat}}=1.15$~GeV/fm$^3$ 
and a hadronic resonance gas equation-of-state (EoS) as before. 
At $T_c$ the hadronic phase starts in chemical equilibrium to  
(approximately) reproduce the measured particle ratios~\cite{BMMRS01},
see above.   
To improve on  previous analyses, the subsequent 
hadronic evolution is now constructed incorporating effective  
meson and (anti-) baryon chemical potentials as in Ref.~\cite{Rapp02}
to preserve the correct (absolute) particle abundances. 

As a second new aspect of the present manuscript, we present an attempt 
to refine the initial conditions of the hydrodynamic evolution. 
More specifically, we will explore ramifications of pre-equilibrium 
collective behavior by introducing appropriate radial velocity 
profiles at the time of complete thermalization. Such effects 
can be associated with pre-thermal re-interactions, a free-streaming
period, or a combination thereof, and turn out to generally improve
the description of transverse momentum spectra of the produced 
particles.

Our article is organized as follows. 
In Sects.~\ref{sec_cent} and \ref{sec_peri} we analyze the 
impact (and interplay) of off-equilibrium hadro-chemistry and modified 
initial collisions on transverse momentum spectra of pions, kaons
and (anti-) protons, both for central and more peripheral collisions 
in comparison to preliminary data at 200~$A$GeV.  
Pertinent predictions for azimuthal anisotropies in non-central collisions
are presented in Sect.~\ref{sec_ell}. We furthermore comment
on implications for the freezeout geometry in Sect.~\ref{sec_fo}, 
and summarize in Sect.~\ref{sec_sum}.   

\section{Particle spectra -- central collisions}
\label{sec_cent}
Let us start by briefly discussing the initial conditions of our 
hydrodynamic calculations. 
According to the $\sim$~15\% larger hadron multiplicity at midrapidity in 
central collisions at 200~$A$GeV~\cite{PHOBOS02,BRAHMS02} 
as compared to 130~$A$GeV, we increase the maximum
entropy-density parameter from $s_0=95$~fm$^{-3}$~\cite{HK02} to 110~fm$^{-3}$
(keeping the equilibration time fixed at $\tau_0=0.6$~fm/$c$ to facilitate the 
interpretation of observed changes).
The correct baryon admixture is obtained by adjusting the entropy-per-baryon to
$S/B=s_0/n_0=250$, constant throughout the 
evolution ($s_0$ and $n_0$ are the initial entropy- and baryon-density 
in the center of the collision, $S$ and $B$ the total entropy and net baryon number). 
The thermodynamic fields in the transverse plane are set to scale 
with a combination of wounded nucleon and 
binary collision profiles as elaborated in Refs.~\cite{HK02,KHHET01},  
which allows for a geometrical prescription to reproduce the multiplicity 
in collisions at finite impact parameter $b$.

The results of our calculations with improved hadro-chemistry are
compared to (preliminary) data for $\pi^-$, $K^-$ and antiproton 
$\pt$-spectra from central Au+Au collisions at 
200~$A$GeV~\cite{spectra200,spectra200cent} in Fig.~\ref{fig:F1}
(the experimental centrality selection of 5~\% is approximated 
by using an average impact parameter $b=2.4$~fm). 
Compared to particle spectra in standard ({\it i.e.}, chemical-equilibrium)
hydrodynamics we find a better description of the overall curved shape
of the hadronic spectra, in particular for low-$\pt$
pions. This is a result of the meson chemical potentials 
($\mu_\pi \approx$~80-100~MeV at freezeout), which amplify the 
Bose-statistics effect. In addition,
the population of heavy resonances also increases after inclusion of
chemical potentials which entails larger contributions at low $\pt$
from their decay products.
At large transverse momenta the hydrodynamic calculations
deviate from the data which is suggestive for
the onset of the hard scattering regime. At exactly which values of 
$\pt$ this occurs, and how this transition depends on the particle 
species, are among the major questions to be clarified. 
{\em E.g.}, high energy partons evolving within a hydrodynamic 
background can be introduced to study the
particle spectra beyond the collective behavior~\cite{HN02}.

As was already observed in Ref.~\cite{HT02}, the expansion of 
the chemically non-equilibrated hadron gas leads to  
slopes for pion spectra that are almost insensitive to the 
decoupling temperature. Proton spectra, on the contrary, clearly favor a 
freezeout at $T\simeq 100$~MeV (thick solid line),     
which corresponds to an energy density $e \simeq 0.075$~GeV/fm$^3$ 
(which is about the same as in previous calculations). 
The thin lines in Fig.~\ref{fig:F1} correspond to decoupling at the phase 
transition (recall that the multiplicity of the individual particle 
species is independent of freezeout due to the chemical potentials).
The experimental pion spectra in the 1-2~GeV range appear flatter than
what follows from the flow generated by hydrodynamic expansion 
with our given initial configuration (at transverse momenta 
$\pt\ge 2$~GeV this is conceivably due to additional perturbative
hard scattering contributions).
To a lesser extent, this is also true for the heavier kaons and protons, 
even at the low freezeout temperature of 100~MeV.  

%
 \begin{figure}[t,b,p]
 \epsfig{file=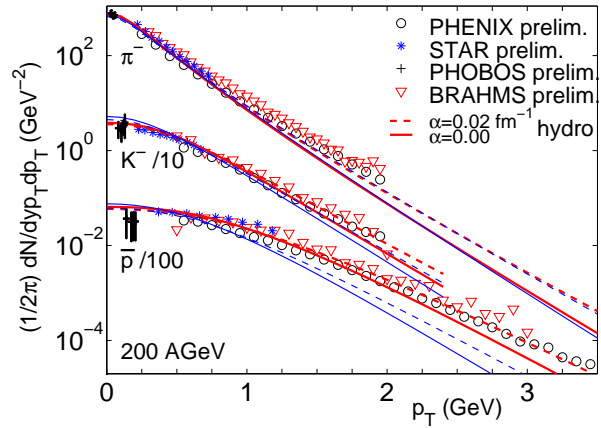,width=8cm}
 \caption{
          $\pi^-$, $K^-$ and antiproton spectra for central collisions at 
          200~$A$GeV ($K^-$ and $\bar p$ spectra are scaled by factors of 
           1/10 and 1/100, respectively). 
          The thick lines represent the results for $T_{\rm dec}=100$~MeV, 
          the thin lines for 165~MeV.
          All calculations are for a thermalization time $\tau_0=0.6$~fm/$c$,
          either {\em without} (solid lines) or {\em with} (dashed lines) 
          an initial transverse boost (see text).
          }
 \label{fig:F1}
 \end{figure}
%

The data thus seem to exhibit somewhat stronger collective expansion than 
developed subsequent to an equilibration time of $\tau_0= 0.6$~fm/$c$. 
Additional radial flow could be generated by assuming still shorter 
equilibration times, \eg, $\tau_0=0.2$~fm/$c$~\cite{ENRR02}. 
It is, however, hard to imagine that particles 
are `born' into thermal equilibrium without allowing for some relaxation time
with rescattering. But even the other extreme, \ie, a period of free streaming, 
induces a non-vanishing radial velocity profile due to a separation of originally 
random particle velocities~\cite{Kolbthesis,HK02}. A realistic situation is 
probably in between the two extremes, essentially pre-equilibrium in character  
with associated rather complicated structures of the generated flow-field 
and energy-density distributions (more exotic phenomena such 
as sphaleron explosions~\cite{Shuryak02} could also play a role). 
As an exploratory study, we here introduce a simplistic initial `seed' 
transverse velocity according to $v_{\rm T}(r)=\tanh ( \alpha \, r )$,
where $r$ is the radial distance from the origin,
superimposed on the original fields at $\tau_0=0.6$~fm/$c$. 
For a value of $\alpha=0.02$~fm$^{-1}$ the initial velocity field for
$r_\perp \le 6$~fm/c is similar in magnitude (although less parabolic) 
to both ($i$) starting the hydrodynamic evolution at earlier time 
($\tau_0=0.2$~fm/$c$ as in Ref.~\cite{ENRR02}) and evolving it to 
$\tau_0=0.6$~fm/$c$, as well as ($ii$) free streaming from 
$\tau=0.2$ to 0.6~fm/$c$.   
The essential difference between ($i$) and ($ii$) lies in the azimuthal
distribution at $\tau_0=0.6$~fm/$c$, to which we will come back to in 
Sect.~\ref{sec_ell}.  It should also be noted that stronger transverse 
flow due to larger transverse
pressure is expected if the longitudinal expansion is not fully
thermalized~\cite{HW02}.

The results with our simple ansatz are represented by the dashed lines in 
Fig.~\ref{fig:F1}, and are found to improve the agreement with experiment, 
up to $\pt\simeq$~2(3.5)~GeV for pions and kaons (antiprotons).
We note that when increasing $\alpha$ to 0.05, the proton spectra
become much flatter than experimentally observed.     
%

\section{Particle spectra -- non-central collisions}
\label{sec_peri}
%
%
 \begin{figure}[!htb]
 \epsfig{file=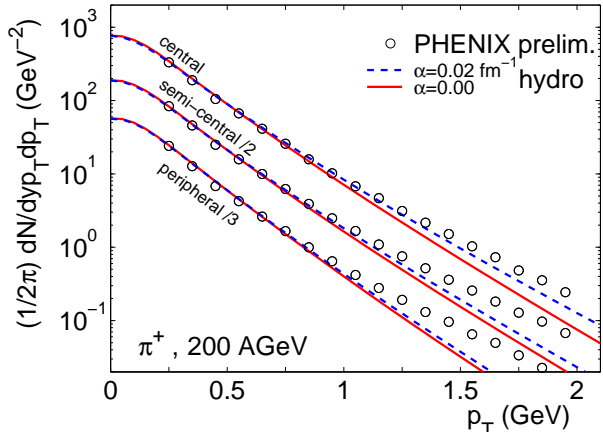,width=8cm}
 \caption{Centrality dependence of positive pion spectra at mid-rapidity
         ($y=0$) in terms of central, semi-central (scaled by 1/2)
          and peripheral collisions (scaled by 1/3).}
 \label{fig:F2}
 \end{figure}
%
%
In Fig.~\ref{fig:F2} we compare preliminary spectra of positive 
pions~\cite{spectra200} to our hydrodynamic results 
at $T_{\rm dec}=100$~MeV in 3 different centrality bins 
    (`central',      $b=2.4$~fm, $N_{\rm part}=343.8$; 
     `semi-central', $b=7  $~fm, $N_{\rm part}=170.8$;
     `peripheral',   $b=9.6$~fm, $N_{\rm part}= 76.6$). 
Again, we display calculations with an initial transverse boost by 
dashed lines.  
As expected, the prerequisites for a hydrodynamic approach (strong 
rescattering and a sufficiently large system size) are increasingly 
invalidated at large impact parameters, reflected by an onset of deviations 
from experiment at smaller transverse momenta (higher-momentum particles can 
rapidly escape the fireball without thermalizing).   
For peripheral collisions the agreement between theory and experiment holds 
for $\pt\le 1$~GeV, which, nevertheless, still 
accounts for more than 96\% of the emitted particles.

Fig.~\ref{fig:F3} shows experimental~\cite{spectra200} and calculated 
proton spectra which are of particular interest in the present context as 
they acquire the largest chemical potentials (\eg, around thermal freezeout 
$\mu_{\rm N}=380$~MeV and $\mu_{\bar {\rm N}}=343$~MeV implying 
$\mu_{\bar{\rm N}}^{{\rm eff}}=723$~MeV, which yields an  antiproton-to-proton 
ratio of 0.72 consistent with experiment~\cite{ppbarratio200}), and are most 
sensitive to collective expansion. 
We find good agreement of theory and experiment at a freezeout
temperature of 100~MeV up to $\pt\simeq$~3.5~GeV
in the central, but only up to $\sim$~2~GeV in the peripheral sample.  
The additional transverse `kick' in the initial state as described above
(dashed lines) is particularly significant for central collisions.  
%
 \begin{figure}[h,t,b,p]
 \epsfig{file=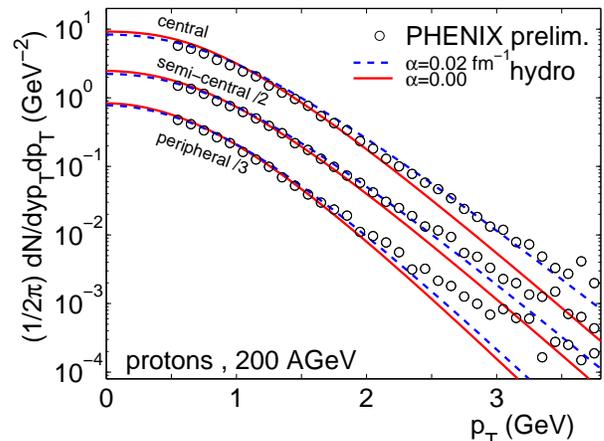,width=8cm}
 \caption{Midrapidity proton spectra for central, 
          semi-cent\-ral (scaled by 1/2) 
          and peripheral collisions (scaled by 1/3).}
 \label{fig:F3}
 \end{figure}
%

Despite the fact that particle densities in the later hadronic stage of the 
expansion are moderate, we conclude that rescattering is strong enough
to allow for a hydrodynamic description until thermal decoupling. 
The correct chemical composition of the hadronic gas is maintained by 
the generation of large chemical potentials, which (at given temperature) 
provide an increased number of scattering partners with larger cross sections 
as compared to a chemically equilibrated environment.
 
We have focused here on positively charged pions and protons.
The corresponding results for $\pi^-$, $K^+$, $K^-$ and $\bar p$ in
non-central collisions are of similar quality. 
The multiplicities and mean transverse momenta of these particles 
are collected in Table 1.

%
\begin{center}
\vspace*{2mm}
 \begin{tabular}{|cc|ccc|ccc|} \hline
            &               &     &$\alpha=0.00$&   &  
\multicolumn{3}{c|} {\hspace*{2mm}$\alpha=0.02$ fm$^{-1}$ }  \\
            &               &   1   &  2    &  3    & 1     & 
\hspace{4mm} 2 \hspace{4mm}   & 3    \\ \hline
            &$\frac{dN}{dy}$& 280.8 & 134.3 & 57.84 & 282.3 & 134.6 & 57.82\\
\raisebox{1.5ex}[-1.5ex] {$\pi^+$}
         &$\la \pt\ra$& 0.398 & 0.392 & 0.375 & 0.419 & 0.405 & 0.383\\ \hline
            &$\frac{dN}{dy}$& 50.18 & 23.99 & 10.33 & 50.43 & 24.05 & 10.33\\
\raisebox{1.5ex}[-1.5ex] {$K^+$}
         &$\la \pt\ra$& 0.619 & 0.608 & 0.572 & 0.660 & 0.634 & 0.589\\ \hline
            &$\frac{dN}{dy}$& 28.08 & 13.44 & 5.798 & 28.13 & 13.44 & 5.794\\
\raisebox{1.5ex}[-1.5ex] {$p$}
         &$\la \pt\ra$& 0.880 & 0.861 & 0.802 & 0.949 & 0.906 & 0.831\\ \hline
 \end{tabular}
\end{center}
{\small   Table 1: Multiplicities and mean transverse momenta (in GeV) 
          of different particles for 3 centrality selections 
          (1=central, 3=peripheral) at $y=0$. The $\bar p$/$p$ ratio 
          is 0.72; $\la \pt\ra$ of antiprotons is 
          within 1\% of the proton value. 
        }
\vspace*{3mm}

\section{Elliptic flow}
\label{sec_ell}
For the same impact parameters as considered above we proceed by studying 
the azimuthal anisotropies of particle spectra~\cite{Ollitrault92}, 
{\it i.e.}, the momentum dependence of elliptic flow as defined by  
$v_2(\pt;\,b)= \la \cos (2\phi) \ra$, 
where the average is taken over the angular distribution of  
particles,  $dN/dy \pt d\pt d\phi$.

Flow anisotropy is generated during the earliest stages of the collision, 
at which the spatial eccentricity of the thermodynamic fields
and the anisotropies in the pressure gradients are the largest. 
The matter is set into anisotropic motion as larger forces are acting 
along the `short' radius of the initial (overlap) ellipse. This motion 
rapidly reduces the spatial anisotropies, thereby bringing further 
generation of momentum anisotropy (\ie, $v_2$) to a stall~\cite{Sorge97,KSH00}. 
If the system evolves in chemical equilibrium, the dominant particle 
species at freezeout are pions, which carry  
the generated anisotropy in their momentum distribution. Their 
differential elliptic flow, $v_2(\pt)$, is then almost   
independent of the decoupling temperature $T_{\rm dec}$~\cite{hydrov2}.
Heavier particles, on the other hand, do exhibit some dependence on 
$T_{\rm dec}$, mainly because of the continuously increasing radial flow 
which shifts the generated anisotropy towards larger transverse momenta.
In the presence of effective chemical potentials the 
contribution of protons to the {\em total} anisotropic flow 
(of all particles) is still small; however, the contribution
of their number to the particle yield is more significant. 
The anisotropic flow  must thus be absorbed by the pions (which, due
to their small masses, adjust their momentum distribution easier).
Through this effect their elliptic flow now also becomes
sensitive to the decoupling temperature, as found in Ref.~\cite{HT02}. 
In addition, the influence of resonance decays is enhanced in the 
chemical off-equilibrium formulation. The heavy resonances, which at 
large transverse momentum carry rather large elliptic flow, decay and 
transfer their elliptic flow to pions at relatively 
low transverse momentum.

In Fig.~\ref{fig:F4} we show results for elliptic flow of pions 
(left panel) and protons (right panel) from the hydrodynamic 
calculation under inclusion of chemical potentials. 
The initial transverse boost as defined in Sect.~\ref{sec_cent}
shifts the anisotropy to larger transverse momenta which implies a 
reduction of $v_2$ at given $\pt$. 
The development of anisotropic flow is additionally hindered 
since it has to form on top of the {\em isotropic} initial boost 
field which we have employed here. The value $\epsilon_p$ at which 
the anisotropy of the energy-momentum tensor of the fluid $T^{\mu \nu}$ 
saturates during the evolution~\cite{KSH00}, is about 25\% smaller 
than without the initial `kick'.  

It is instructive to compare these results to the scenario
where the equilibration time is set to very small values 
($\tau_0=0.2$~fm/$c$ for the dashed-dotted curves in Fig.~\ref{fig:F4}).
As elucidated in Sect.~\ref{sec_cent}, this generates as much {\em radial}
flow as the superimposed profile at $\tau_0=$~0.6~fm/$c$ does.
However, the elliptic flow is larger than in the former case, but
{\em not} significantly different from using an equilibration time
of 0.6~fm/$c$ {\em without} initial kick. This is due to the fact
that, without initial kick, $v_2$ saturates for either equilibration 
time at approximately the same value~\footnote{Note that this is no longer
true for significantly larger $\tau_0$, \eg~2~fm/$c$, 
for which $v_2$ is significantly reduced and underpredicts the data
already at 130~$A$GeV}.

%
 \begin{figure}[t,b,p]
 \epsfig{file=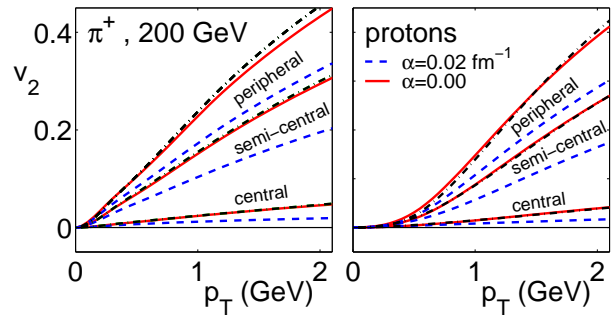,width=8cm}
 \caption{Elliptic flow of positively charged pions (left) and protons (right)
          for three different impact parameters.
          The dashed lines include an initial transverse boost as described 
          in the text. Dashed-dotted lines represent the results when 
          assuming thermalization at $\tau_0=0.2\;{\rm fm}/c$ with $\alpha=0$.}
 \label{fig:F4}
 \end{figure}
%

The experimentally observed elliptic flow reaches a 
limiting maximal value as a function of transverse momentum.
The PHENIX collaboration has pointed out~\cite{spectra200} that 
this saturation is reached at smaller transverse momenta for pions 
than for protons, and that the saturation value appears to be larger
for the latter.
Within the hydrodynamic framework this reflects the earlier breakdown of
the strong rescattering assumption for pions, which for protons 
remains valid up to higher $\pt$ due to larger (average) 
scattering cross sections ($\bar \sigma_{\pi N} > \bar \sigma_{\pi\pi}$).  
This is corroborated by the description of the single-particle spectra which 
extends to larger $\pt$ for (anti-) protons than for pions.
The different $v_2$-saturation momenta for mesonic and baryonic elliptic flow
are also consistent with the formation of hadrons via 
quark-coalescence~\cite{Voloshin02}. Within this picture one similarly 
expects a larger saturation value of $v_2$ for protons than for pions.

\section{Freeze-out geometry}
\label{sec_fo}
Let us finally comment on the implications of our results 
for the freezeout geometry of the hadronic system.  
In Ref.~\cite{Teaney02} it was pointed out that the relation between 
energy density and pressure, $e(p)$, for the hadronic equation of state 
is barely modified by the introduction of chemical potentials. 
Therefore, the space-time evolution of the system, 
which is largely driven by this relation, 
is not substantially altered either. 
A large change, however, occurs in the relation
between temperature and energy density, $T(e)$, 
which thus influences the construction of the 
freezeout hypersurface and the thermal properties of the fluid on this 
surface.  {\it E.g.}, in chemical equilibrium the energy density at $T=130$~MeV
corresponds to a temperature of only 100~MeV in the presence of large chemical
potentials, since the latter increase particle and energy 
densities approximately by pertinent fugacity factors ${\rm e}^{\mu /T}$.  
Therefore, the freezeout hypersurface of the hydrodynamic calculations 
in chemical off-equilibrium is not much different from the 
hypersurface of previous calculations if freezeout is performed 
at a comparable energy density (\ie, the freezeout temperature is adapted 
accordingly). In both cases, the fireball decouples at about 15~fm/$c$ 
after equilibration (in central collisions) 
and has about the same spatial extent. 
Only after inclusion of the initial radial flow profile is the lifetime 
shortened by $\sim$~15\%, and the transverse expansion increases by
about the same percentage. For observables, this entails smaller
longitudinal correlation radii (which reflect the system's lifetime) 
but only slightly larger sideward radii. This effect reduces   
the discrepancies between calculated and measured 
Hanbury-Brown and Twiss (HBT) radii by a few 
percent~\cite{HK02}, but is not sufficient by itself. 
Additional effects, such as viscosity~\cite{viscosity02}, 
large partonic cross sections in the early phases~\cite{LKP02}, 
or a refined treatment of hadronic rescattering~\cite{SBD01} and 
freezeout~\cite{SBHP02} (including, \eg,  a large $\rho$-meson 
width as predicted in Ref.~\cite{Ra02}), seem to be required to 
fully resolve the "HBT puzzle".

\section{Summary}
\label{sec_sum}
Based on a resonance gas equation of state which explicitly 
incorporates hadrochemical freezeout by employing chemical potentials 
for (stable) mesons and baryons in the hadronic evolution, we have 
performed hydrodynamic simulations of heavy ion-collisions at full 
RHIC energy.  
We have compared the results for pion, kaon, and proton $\pt$-spectra to 
preliminary data from 200~$A$GeV Au+Au collisions at 
different centralities. Our 
investigations indicate the necessity of an initial (pre-hydrodynamic) 
transverse flow to better account for the slopes of the observed 
spectra. 
Good agreement with preliminary data for transverse momentum spectra 
in central collisions is obtained up to $\sim$~1.5-2~GeV for pions, and 
up to at least 3~GeV for protons.  
We further studied the influence of hadrochemistry and initial flow 
on elliptic flow and source geometry. The former has been presented as 
a prediction for pions and protons for upcoming experimental analyses.
For the latter, some improvement with respect to the discrepancy
between model and data has been found, but additional effects remain
mandatory.  

\vspace{0.3cm}


{\bf Acknowledgments:} \\
We thank U. Heinz and E.V. Shuryak for  fruitful discussions and 
critical remarks on the manuscript. 
This work was supported in part by the U.S. Department of Energy under
grant No. DE-FG02-88ER40388.
PFK acknowledges support from the Alexander von Humboldt Foundation
through a Feodor-Lynen Fellowship.


\end{narrowtext}

\end{document}